\documentclass[graybox]{svmult}

\usepackage{mathptmx}       
\usepackage{helvet}         
\usepackage{courier}        
\usepackage{type1cm}        
\usepackage{makeidx}         
\usepackage{graphicx}        
\usepackage{multicol}        
\usepackage[bottom]{footmisc}

\usepackage{epsfig,amsmath,hyperref}
\usepackage{amsfonts,amssymb,amsmath,amsopn,amsxtra}
\usepackage{graphicx}
\usepackage{rotating}

\newcommand{\R}{\mathbb{R}}

\newcommand{\e}{\mathrm{e}}

\makeindex             

\begin{document}

\title*{Schr\"odinger operators with a switching effect}
\author{Pavel Exner}
\institute{Pavel Exner \at Nuclear Physics Institute, Czech Academy of Sciences, Hlavn\'{i} 130, 25068 \v{R}e\v{z} near Prague, Czechia, \email{exner@ujf.cas.cz}}
%
%
\maketitle

\abstract{This paper summarizes the contents of a plenary talk given at the 14th Biennial Conference of Indian SIAM in Amritsar in February 2018. We discuss here the effect of an abrupt spectral change for some classes of Schr\"odinger operators depending on the value of the coupling constant, from below bounded and partly or fully discrete, to the continuous one covering the whole real axis. A prototype of such a behavior can be found in
Smilansky-Solomyak model devised to illustrate that an an irreversible behavior is possible even if the heat bath to which the systems is coupled has a finite number of degrees of freedom and analyze several modifications of this model, with regular potentials or a magnetic field, as well as another system in which $x^py^p$ potential is amended by a negative radially symmetric term. Finally, we also discuss resonance effects in such models.}

\section{Introduction}
\label{sec:intro}

The class of problems we are going to discuss here has a twofold motivation. Let us start with physics. It is well known that while the equations of motion governing quantum dynamics are invariant with respect to time reversal, we often encounter quantum systems behaving in an irreversible way, for instance, spontaneous decays of particles and nuclei, inelastic scattering processes in nuclear, atomic or molecular systems, or the current passing through a microscopic element attached to poles of a battery. Furthermore, an irreversible process \emph{par excellence} is, of course, the \emph{wave packet reduction} which is the core of Copenhagen description of a measuring process performed on a quantum system.

The description of such a process is typically associated with enlarging the state Hilbert space, conventionally referred to as coupling the system to a \emph{heat bath}. It is generally accepted that to obtain an irreversible behavior through such a coupling, the model has to exhibit typical properties, in particular
\begin{itemize}
 \setlength{\itemsep}{0pt}
 \item the bath is a system with infinite number of degrees of freedom
 \item the bath Hamiltonian has a continuous spectrum
 \item the presence (or absence) of irreversible modes is determined by the energies involved rather than the coupling strength
\end{itemize}
While this all is without any doubt true in many cases, one of our aims here is to show that \emph{neither of the above need not be true in general}. To make this point, Uzy Smilansky constructed a simple model which will be our starting point here. In a sense he did a similar thing as Agatha Christie: when some people tried to introduce in the 1920s mystery rules saying, in particular, that in any such such book there must be a single murderer, she wrote \emph{Murder on the Orient Express} in which everyone is a killer, except Hercule Poirot, of course.

On the other hand, as a mathematician one may ask whether a \emph{small change of the coupling constant} can have a profound influence on the spectrum. Posed like that the answer is trivial: consider the the one-dimensional Schr\"odinger operator
$$
H_\lambda = -\frac{\mathrm{d}^2}{\mathrm{d}x^2} + \lambda x^2\,;
$$
it is obvious that all $\lambda=\omega^2>0$ the spectrum of such an operator is \emph{purely discrete}, $\sigma(H_\lambda) = \{(2n+1)\omega:\: n=0,1,\dots\}$,  while for $\lambda=0$ and $\lambda<0$ we have $\sigma(H_\lambda) = [0,\infty)$ and $\sigma(H_\lambda) = \mathbb{R}$, respectively.
A much more subtle question is whether similar things could happen if the potential modification concerns a \emph{small part} of the configuration space, or even a \emph{`set of zero measure'}. Smilansky model and its various modifications we are going to discuss provide an \emph{affirmative answer}.

Let us describe briefly the contents of the paper. In the next section we summarize the known results about Smilansky-Solomyak model and present a numerical method to analyze its discrete spectrum. Sec.~\ref{sec:variations} is devoted to discussion of various modifications of the model consisting, in particular, in replacing the $\delta$ interaction `channel' by a regular potential one or, on the contrary, by the more singular $\delta'$ interaction, or adding a homogeneous magnetic field. In Sec.~\ref{sec:another} we discuss another model exhibiting a similar behaviour, a two-dimensional Schr\"odinger operator with the potential consisting of the $x^py^p$ part amended by a negative radially symmetric term. In Sec.~\ref{sec:resonances} we return to the original Smilansky-Solomyak model and show that it also has a rich resonance structure. Finally, in conclusion we will mention several open questions.

\section{Smilansky-Solomyak model}
\label{sec:smisom}

Let us first describe the model proposed by Uzy Smilansky in \cite{Sm04} which in its simplest form describes a one-dimensional system interacting with a caricature heat bath represented by a single harmonic oscillator. Its mathematical properties and various extensions were subsequently analyzed by Michael Solomyak -- let us pay a memory to this great mathematician who left us two years ago -- and coauthors in \cite{ES05I, ES05II, NS06, RS07, S04FAA, S06, S06JPA} from the spectral point view, the corresponding time evolution was discussed in \cite{G11, G18}.

With this history of the problem in mind, it is appropriate to speak of the \emph{Smilansky-Solomyak model}. At the same time, it is useful to note that while mathematically it is the same thing, physically there may be two ways in which the system is understood. In the original Smilansky treatment one considers two one-dimensional systems coupled mutually, while Solomyak et al. interpreted it in PDE terms as being described by a two-dimensional Schr\"odinger operator,
\begin{equation} \label{Hsmil}
H_\mathrm{Sm}=-\frac{\partial^2}{\partial x^2}
+\frac12\left( -\frac{\partial^2}{\partial y^2}+y^2 \right)
+\lambda y\delta(x),
\end{equation}
on $L^2(\R^2)$; it is easy to see that that one may consider $\lambda\ge 0$ only without loss of generality. We will stick here to the latter interpretation because it opens way to a wider class of possible generalizations.

Let us summarize the known results about spectral properties of the operator \eqref{Hsmil}:
\begin{itemize}
 \setlength{\itemsep}{2pt}

\item The existence of a \emph{spectral transition:} if $|\lambda|>\sqrt{2}$ the particle can escape to infinity along the singular `channel' in the $y$ direction. In spectral terms, this corresponds to the switch from a positive spectrum to a below unbounded one at $|\lambda|=\sqrt{2}$. At the heuristic level, the \emph{mechanism} of this spectral transition is easy to understand: we have an effective variable decoupling far from the $x$-axis and the oscillator potential competes there with the $\delta$ interaction eigenvalue $-\frac14 \lambda^2y^2$.

\item The \emph{eigenvalue absence:} for any $\lambda\ge 0$ there are \emph{no eigenvalues $\ge \frac12$}. If $|\lambda|>\sqrt{2}$, the point spectrum of $H_\mathrm{Sm}$ is \emph{empty}.

\item The \emph{existence of eigenvalues:} in the subcritical case, $0<|\lambda|<\sqrt{2}$, we have $H_\mathrm{Sm}\ge 0$. The point spectrum is then nonempty and finite, and
\begin{equation} \label{Solasympt}
N(\textstyle{\frac12},H_\mathrm{Sm}) \sim \frac{1}{4\sqrt{2(\mu(\lambda)-1)}}
\end{equation}
holds as $\lambda\to\sqrt{2}-$, where $\mu(\lambda):= \sqrt{2}/\lambda$.

\item The \emph{absolute continuity:} in the supercritical case, $|\lambda|>\sqrt{2}$, we have $\sigma(H_\mathrm{Sm}) = \sigma_\mathrm{ac}(H_\mathrm{Sm})=\R$.

\end{itemize}

We are not going to give proofs of these claims referring to the papers quoted above, instead we will show how the the discrete spectrum can be found \emph{numerically} following \cite{ELT17a} which can provide additional insights. At the time, however, the method we use, rephrasing the task as a \emph{spectral problem for Jacobi matrices} is the core of the proofs done by Solomyak et al. providing thus a feeling of what is the technique involved.

\begin{figure}
\vspace{-1em}
\begin{center}
\hspace{0em}\includegraphics[angle=0,width=8cm]{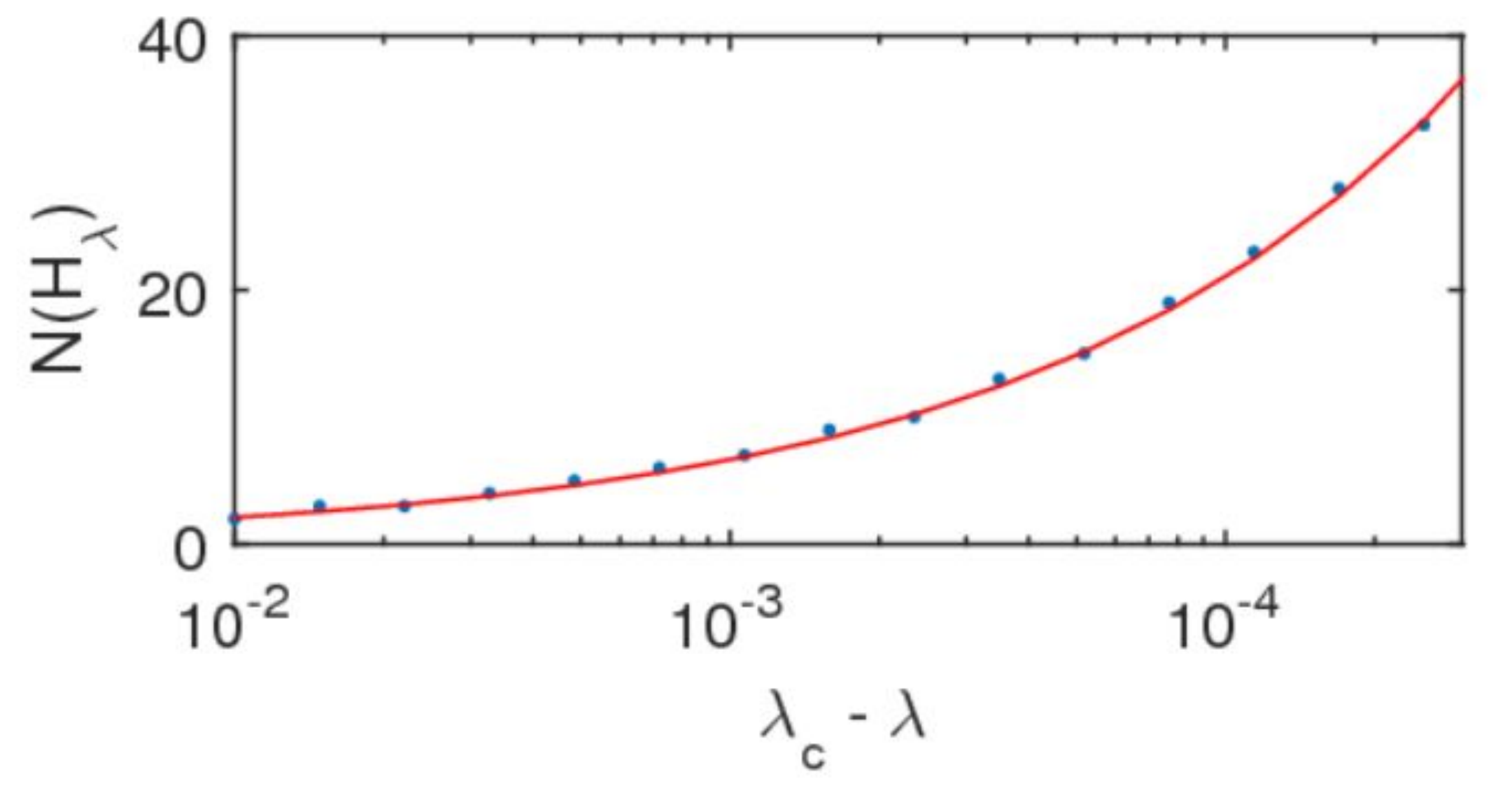}
\caption{Number of eigenvalues of $H_\mathrm{Sm}$ \emph{vs.} the asymptotics \eqref{Solasympt}}
\label{fig:3}
\end{center}
\end{figure}

\begin{figure}
\vspace{-1em}
\begin{center}
\includegraphics[angle=0,width=8cm]{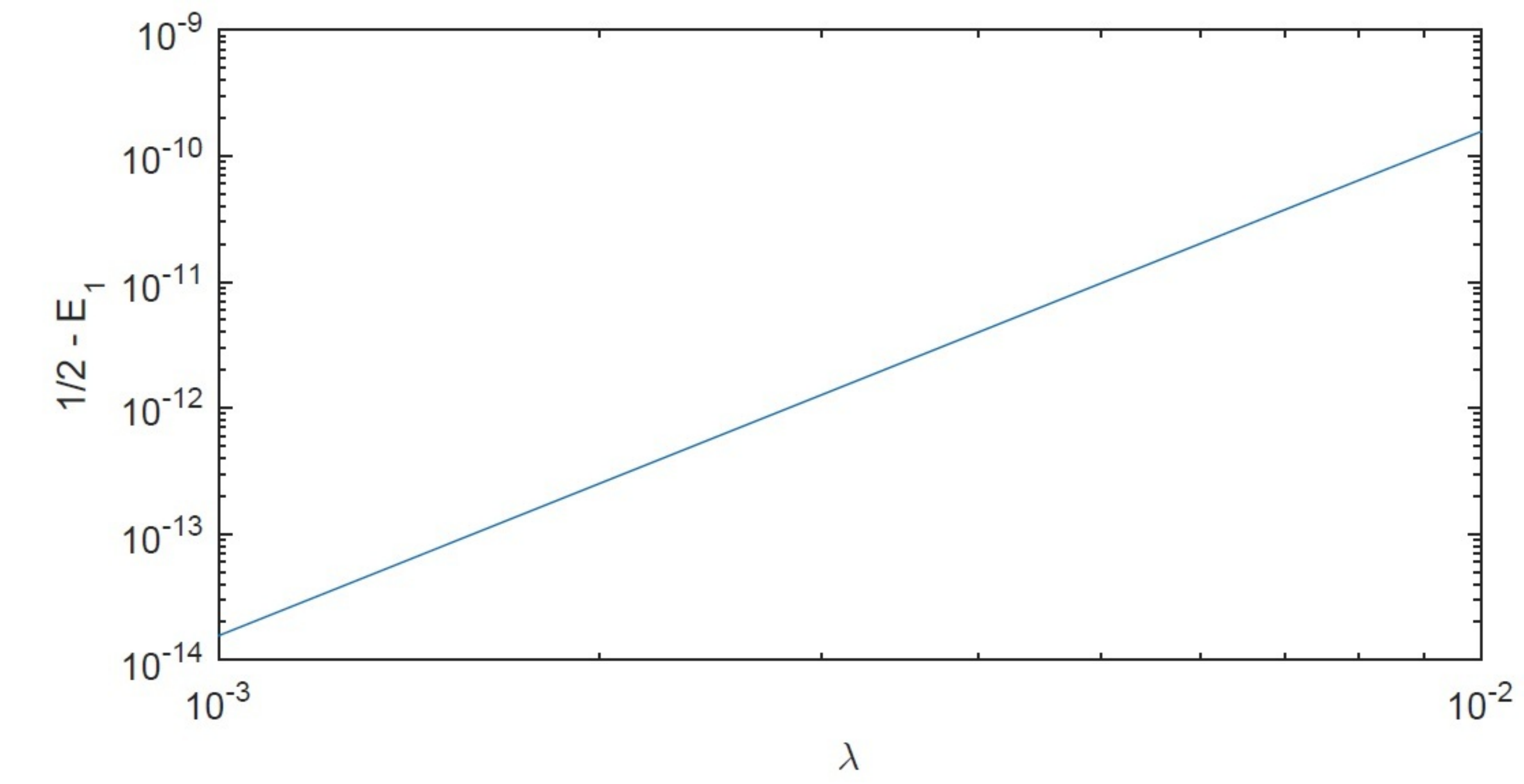}
\caption{Weak coupling asymptotics of $H_\mathrm{Sm}$}
\label{fig:4}
\end{center}
\end{figure}

In the halfplanes $\pm x>0$ the wave functions can be expanded using the `transverse' base spanned by the functions
\begin{equation} \label{HObase}
\psi_n(y) = \frac{1}{\sqrt{2^n n! \sqrt\pi}}\: \mathrm{e}^{-y^2/2} H_n(y)
\end{equation}
corresponding to the oscillator eigenvalues $n+\frac12,\: n=0,1,2,\dots\,$.
Furthermore, one can make use of the mirror symmetry w.r.t. \mbox{$x=0$} and divide $H_\lambda$ into the trivial odd part $H_\lambda^{(-)}$ and the even part $H_\lambda^{(+)}$ which is equivalent to the operator on the halfplane, $L^2(\mathbb{R}\times(0,\infty))$, with the same symbol determined by the boundary condition
\begin{equation} \label{halfplane}
f_x(0+,y) = \frac12\, \alpha y f(0+,y).
\end{equation}
We substitute the Ansatz
\begin{equation} \label{Ansatz}
f(x,y) = \sum_{n=0}^\infty c_n\, \mathrm{e}^{-\kappa_n x} \psi_n(y)
\end{equation}
with $\kappa_n:= \sqrt{n+\frac12-\epsilon}$ into \eqref{halfplane}; this yields for the sought solution with the energy $\epsilon$ the equation
\begin{equation} \label{secular}
B_\lambda c = 0,
\end{equation}
where $c$ is the coefficient vector and $B_\lambda$ is the operator in $\ell^2$ with the representation
\begin{equation} \label{Jacobi}
(B_\lambda)_{m,n} = \kappa_n \delta_{m,n} + \frac12\lambda (\psi_m,y \psi_n).
\end{equation}
It is obvious that the matrix is in fact tridiagonal because
\begin{equation} \label{tridiagonal}
(\psi_m,y \psi_n) = \frac{1}{\sqrt{2}} \big( \sqrt{n+1}\, \delta_{m,n+1} + \sqrt{n}\, \delta_{m,n-1} \big).
\end{equation}
To solve the equation \eqref{secular} numerically one truncates the matrix \eqref{Jacobi}. The size depends on $\lambda$, the most difficult is the weakly bound state corresponding to small $\lambda$ where the truncation size should be of order of $10^4$ to achieve a numerically stable solution.
\begin{figure}
\begin{center}
\includegraphics[angle=0,clip, trim=2cm 16.2cm 10cm 9.3cm, width=0.8\textwidth]{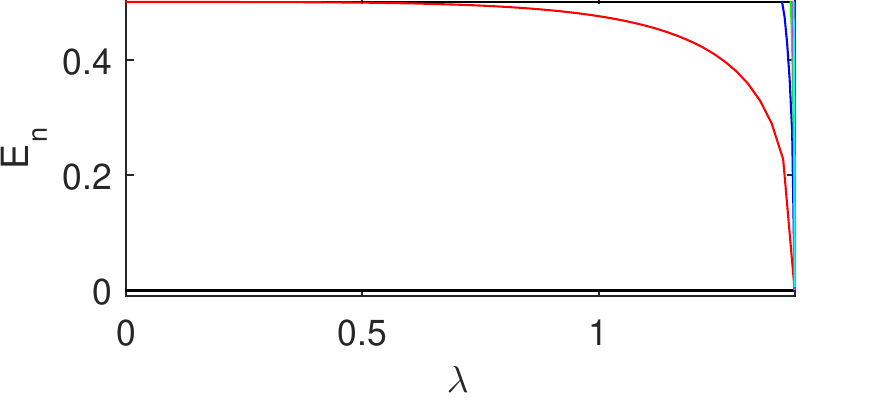}
\caption{Discrete spectrum of $H_\mathrm{Sm}$ as a function of the coupling constant $\lambda$}
\label{fig:1}
\end{center}
\end{figure}
The result is plotted in Fig.~\ref{fig:1}. In coincidence with the theoretical result quoted above the discrete spectrum is nonempty for nonzero $\lambda$. It may seem that it consists of a single eigenvalues but a closer look shows that the second one appears at $\lambda\approx
1.387559$; the next thresholds are $1.405798,1.410138,1.41181626,1.41263669, \dots\:$.
\begin{figure}
\vspace{-.5em}
\begin{center}
\hspace{-2em}\includegraphics[angle=0,width=8cm]{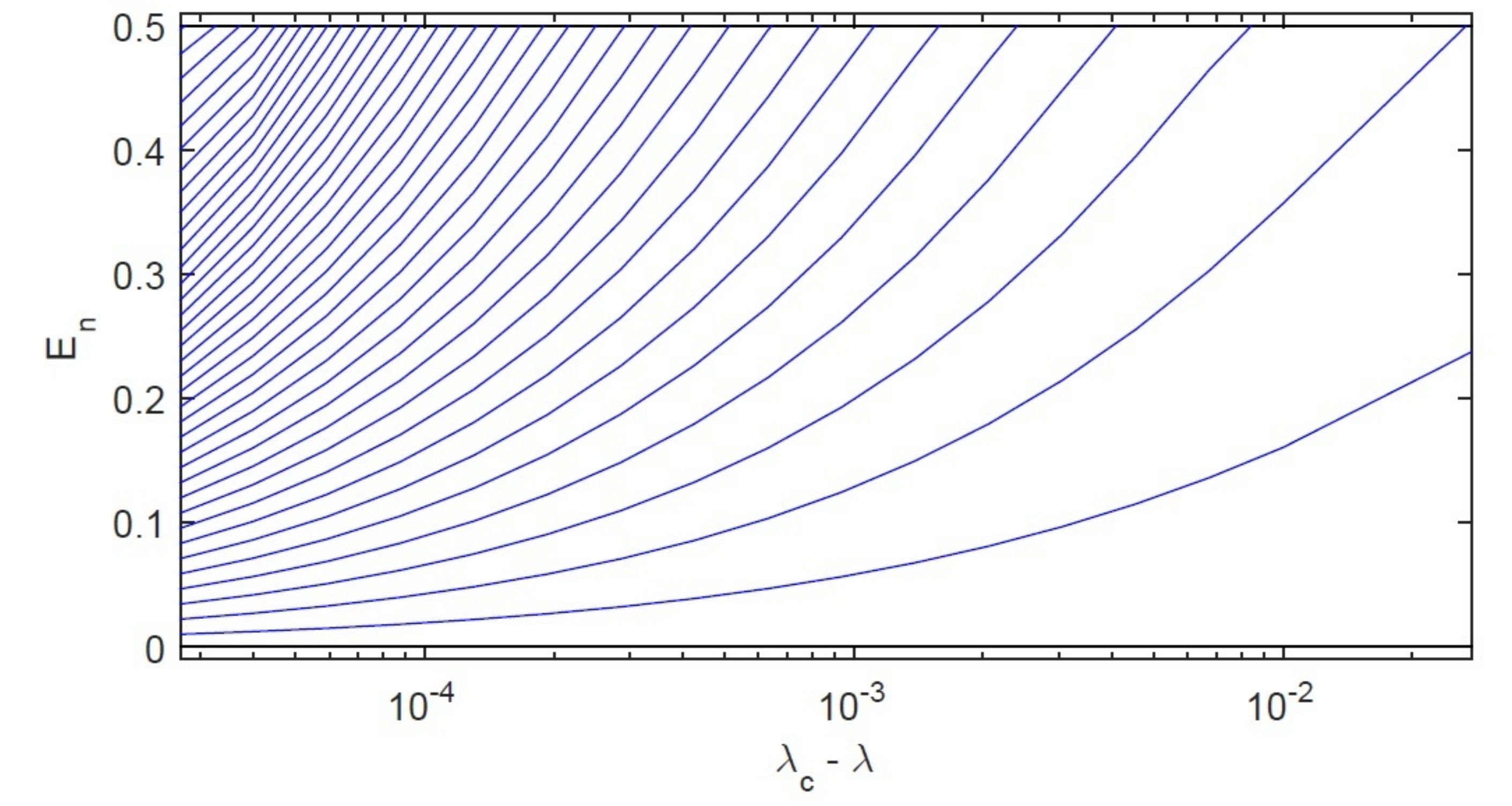}
\caption{Discrete spectrum of $H_\mathrm{Sm}$ near the critical value of the coupling constant}
\label{fig:2}
\end{center}
\end{figure}
To have a better insight one can plot the discrete spectrum near the critical coupling in the semilogarithmic scale as shown in Fig.~\ref{fig:2}.
We see that in this regime many eigenvalues appear which gradually fill the interval $(0,\frac12)$ as the critical value $\lambda=\sqrt{2}$ is
approached. Fig.~\ref{fig:3} shows a comparison of their number indicated by dots with the asymptotics \eqref{Solasympt}; we see a perfect fit.
The numerical solution also indicates other properties. For instance, plotting in Fig.~\ref{fig:4} the eigenvalue curve for small values of $\lambda$ in the logarithmic scale we see that it behaves as $E_1(\lambda) = \frac12 - c\lambda^4 +o(\lambda^4)$ as $\lambda\to 0$, with $c\approx 0.0156$.
In fact, the coefficient value can be found exactly to be $c=0.015625$. To this aim, we write the equation \eqref{secular} explicitly in components as
\begin{eqnarray}
    \sqrt{\mu_\lambda}c^\lambda_0 + \frac{\lambda}{2\sqrt{2}}c^\lambda_1 = 0,
& \phantom{A} \nonumber \\[-.3em] \label{component} \\[-.3em]
    \frac{\sqrt{k}\lambda}{2\sqrt{2}}c^\lambda_{k-1} + \sqrt{k + \mu_\lambda}c^\lambda_k
      + \frac{\sqrt{k+1}\lambda}{2\sqrt{2}}c^\lambda_{k+1} = 0, \nonumber
& \;\; k\ge 1\,,
\end{eqnarray}
where $\mu_\lambda:= \frac12 - E_1(\lambda)$ and $c^\lambda= \{c^\lambda_0,c^\lambda_1,\dots\}$ is the corresponding normalized eigenvector of $B_\lambda$. Using the above relations and simple estimates, we get from here
\begin{equation} \label{component2}
\sum_{k=1}^\infty |c_k^\lambda|^2 \le \frac34 \lambda^2 \quad \text{and} \quad c_0^\lambda = 1+\mathcal{O}(\lambda^2)
\end{equation}
as $\lambda\to 0+$, hence we have in particular $c_1^\lambda = \frac{\lambda}{2\sqrt{2}} +\mathcal{O}(\lambda^2)$. The first of the above relations then gives $\mu_\lambda = \frac{\lambda^4}{64} + \mathcal{O}(\lambda^5)$ as $\lambda\to 0+$, in other words
\begin{equation} \label{Smilweak}
E_1(\lambda) = \frac12 - \frac{\lambda^4}{64} + \mathcal{O}(\lambda^5),
\end{equation}
however, the mentioned coefficient $0.015625$ is nothing else than $\frac{1}{64}$. Furthermore with the knowledge of the solution to \eqref{secular} we can return to \eqref{Ansatz} and compute the eigenfunctions.
\begin{figure}
\vspace{-8em}
\begin{center}
\includegraphics[angle=0,trim=0cm 5cm 0cm 1cm, width=10cm]{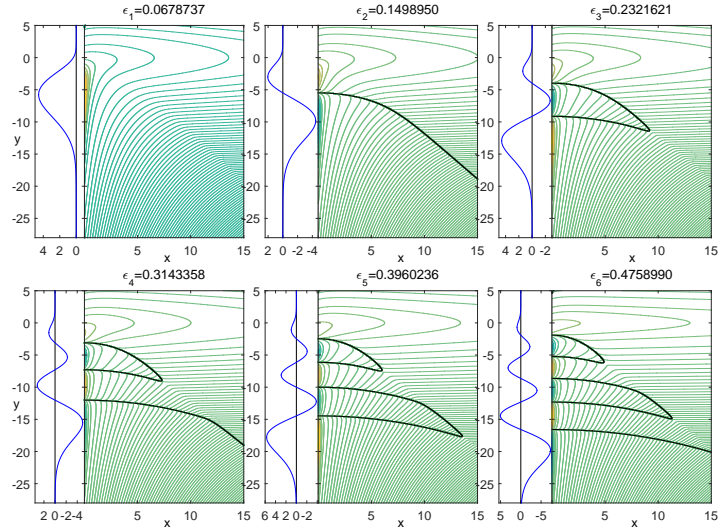}
\caption{The eigenfunctions of $H_\mathrm{Sm}$ for $\lambda=1.4128241$}
\label{fig:5}
\end{center}
\end{figure}
In Fig.~\ref{fig:5} we plot them for a value close to the critical one, namely $\lambda=\sqrt{2}-0.0086105$. As expected, they are stretched along the axis of the oscillator `channel' and the part of the $y$-axis where the singular interaction is attractive; the he curves at the left side show the $y$-cuts of the eigenfunctions. The ground state has no zeros and the number of the nodal lines of the $n$th eigenfunction, $n=1,2,\dots\:$, is $[\frac12 n]$, thus only the first excited state is Courant sharp.

\section{Variations on the model}
\label{sec:variations}

We have mentioned in the introduction that various extensions of the Smilansky-Solomyak model described above had been worked out, for instance, using  a `heat bath' consisting of more than one oscillator, replacing the line by a loop or a graph, etc. We will not discuss them, instead we will analyze several other modifications.

\subsection{Regular version of the model}
\label{subsec:regular}

The first one is motivated by the question whether one can observe a similar effect for Schr\"odinger operators with the $\delta$ interaction replaced by a regular potentials. It was asked by Italo Guarneri in \cite{G11} with a clear motivation: he employed (a modification of) the system described above as a model of the \emph{measuring process} in quantum mechanics in which the supercritical behaviour is interpreted as a wave packet reduction. This naturally inspires the question how the corresponding classical dynamics would look like, and this in turn requires a setting in which the problem can be analyzed in terms of classical mechanics; the first step in this direction has been made in the recent paper \cite{G18}.

We observe first that the coupling cannot be now linear in $y$ and the profile of the channel has to change with the variable $y$. We replace the $\delta$ by a \emph{family of shrinking potentials} the mean of which matches the $\delta$ coupling constant, $\int U(x,y)\,\mathrm{d}x
\sim y$. This can be achieved, e.g., by choosing $U(x,y)=\lambda y^2V(xy)$ for a fixed function $V$. This motivates us to investigate the following operator on $L^2(\mathbb{R}^2)$,
\begin{equation} \label{regular}
H=-\frac{\partial^2}{\partial x^2}-\frac{\partial^2}{\partial y^2}+\omega^2y^2
-\lambda y^2V(xy),
\end{equation}
where $\omega,\,a$ are positive constants and the potential $V$ is a nonnegative function with bounded first derivative and $\mathrm{supp}\,V\subset[-a,a]$. By Faris-Lavine theorem $H$ is e.s.a. on $C_0^\infty (\mathbb{R}^2)\,$ -- cf.~\cite[Thms.~X.28 and X.38]{RS} -- and the same argument can be applied to various generalizations of the operator \eqref{regular}, with more than one `decay' channel, periodicity in the variable $x$, etc.

To state the result we need a one-dimensional comparison operator $L=L_V$, namely
\begin{equation} \label{comparison}
L=-\frac{\mathrm{d}^2}{\mathrm{d}x^2}+\omega^2-\lambda V(x)
\end{equation}
on $L^2(\mathbb{R})$ with the domain $H^2(\mathbb{R})$. What matters is the sign of its spectral threshold; since $V\ge 0$, the latter is a monotonous function of $\lambda$ and there is a $\lambda_\mathrm{crit} >0$ at which the sign changes. First of all, we have the following result \cite{BE14}.

\begin{theorem} \label{thm 2}
Under the stated assumption, the spectrum of the spectrun of $H$ is bounded from below provided the operator $L$ is positive.
\end{theorem}
\emph{Sketch of the proof}.
The claim can proved using \emph{Neumann bracketing}, imposing additional boundary conditions at the lines $y=\pm \ln n$, $n=2,3,\dots$, and showing that the components of $H$ in these strips have a uniform lower bound by an operator unitarily equivalent to $L$, cf. \cite{BE14} for details.
\qed

One the other hand, in the supercritical case when the transverse channel principal eigenvalue dominates over the harmonic oscillator contribution, the spectral behavior changes \cite{BE14}.

\begin{theorem}
Under our hypotheses, $\sigma(H)=\mathbb{R}$ holds if $\inf\sigma(L)<0$.
\end{theorem}
\emph{Sketch of the proof}.
The argument relies on a choice of an appropriate Weyl sequence: we have to find $\{\psi_k\}_{k=1}^\infty\subset D(H)$ such that $\|\psi_k\|=1$ which contains no convergent subsequence, and at same time
\begin{equation} \label{Weyl}
\|H\psi_k-\mu\psi_k\|\to0\quad\text{as}\quad k\to\infty.
\end{equation}
Specifically, we choose
$$
\psi_k(x,y)=h(xy)\,\e^{i\epsilon_\mu(y)} \chi_k\left(\frac{y}{n_k}\right)+
\frac{f(xy)}{y^2}\,\e^{i\epsilon_\mu(y)} \chi_k\left(\frac{y}{n_k}\right),
$$
where $\epsilon_\mu(y):= \int_{\sqrt{|\mu|}}^y \sqrt{t^2+\mu}\,\mathrm{d}t$, $\:h$ is the normalized ground-state eigenfunction of $L$, furthermore $\,f(t):=-\frac{i}{2}\,t^2h(t)$, and finally, $\chi_k$ are suitable, compactly supported mollifier functions, cf. \cite{BE14} for details.
\qed

The regular version shares also other properties with the original Smilansky-Solomyak model, namely \cite{BE17a}:
\begin{itemize}
\setlength{\itemsep}{2pt}

\item in the subcritical case, $\inf\sigma(L)>0$, we have $\sigma_\mathrm{ess}(H) = [\omega,\infty)$ and a nonempty $\sigma_\mathrm{disc}(H) \subset [0,\omega)$

\item in the critical case, $\inf\sigma(L)=0$, we have $\sigma(H) = \sigma_\mathrm{ess}(H) = [0,\infty)$

\end{itemize}

\subsection{Magnetic version of the model}
\label{subsec:magnetic}

One can also consider another modification of Smilansky-Solomyak model in which the system is placed into a \emph{homogeneous magnetic field} perpendicular to the plane representing the configuration space, described thus by the Hamiltonian
\begin{equation} \label{mgSmil}
H=(i\nabla +A)^2+\omega^2 y^2+\lambda y \delta(x),
\end{equation}
where $A$ is a suitable vector potential; note that in this case the original Smilansky interpretation is lost. The spectral properties are similar, in the subcritical case we now have $\sigma_\mathrm{ess}(H(A)) = [\sqrt{B^2+\omega^2},\infty)$ but again a sufficiently small nonzero $\lambda$ gives rise to a discrete spectrum which fills the interval $[0,\sqrt{B^2+\omega^2})$ as $|\lambda|$ approaches the critical coupling $2\omega$, and above this value the spectrum fills the whole real line. The effect of the magnetic field on the regular version of the model is similar, cf.~\cite{BE17b} for details.

\subsection{The $\delta'$ version of the model}
\label{subsec:delta'}

One can also say that the spectral transition effect is robust. To illustrate this claim let us consider the version of the model in which the interaction is replaced by a \emph{more singular} one, specifically the one known as $\delta'\:$ \cite{AGHH}. The Hamiltonian then corresponds to the differential expression
\begin{equation} \label{delta'Smil}
  H_\beta \psi(x,y)= - \frac{\partial^2 \psi}{\partial x^2}(x,y) +\frac{1}{2}\left(- \frac{\partial^2 \psi}{\partial y^2}(x,y) +y^2\psi(x,y)\right)
\end{equation}
with the domain consisting of $\psi\in H^2((0,\infty)\times\mathbb{R})\oplus H^2((-\infty, 0)\times\mathbb{R})$ such that
\begin{equation} \label{delta'domain}
  \psi(0+,y)-\psi(0-,y) = \frac{\beta}{y} \frac{\partial \psi}{\partial x}(0+,y)\,,\quad
  \frac{\partial \psi}{\partial x}(0+,y) = \frac{\partial \psi}{\partial x}(0-,y).
\end{equation}

The problem can be treated by a modification of the methods employed in \cite{ES05I, ES05II, NS06, RS07, S04FAA, S06, S06JPA} leading to the following results which we present referring to \cite{EL18} for the proofs. Let $\mathfrak{m}_\mathrm{ac}$ be the multiplicity of the absolutely continuous spectrum.
\begin{theorem}
The spectrum of operator $H_0$ is purely \emph{ac}, $\sigma(H_0) = [\frac12,\infty)$ with the multiplicity $\mathfrak{m}_\mathrm{ac}(E,H_0) = 2n$ for $E\in(n-\frac12,n+\frac12)$, $\,n\in \mathbb{N}$.  For $\beta>2\sqrt{2}$ the \emph{ac} spectrum of $H_\beta$ coincides with $\sigma(H_0)$. For $\beta \leq 2\sqrt{2}$ there is a new branch of continuous spectrum added to the spectrum; for $\beta = 2\sqrt{2}$ we have $\sigma(H_\beta) = [0,\infty)$ and for $\beta < 2\sqrt{2}$ the spectrum \emph{covers the whole real line}.
\end{theorem}
We note in passing that with the standard convention used here, \emph{small} values of the parameter $\beta$ represent a \emph{strong} coupling.

\begin{theorem}
Assume $\beta\in (2\sqrt{2},\infty)$, then the discrete spectrum of $H_\beta$ is \emph{nonempty} and lies in the interval $(0,\frac12)$. The number of eigenvalues is approximately given by
$$
  \frac{1}{4\sqrt{2\left(\frac{\beta}{2\sqrt2} - 1\right)}}\quad \mathrm{as}\quad \beta\to2\sqrt{2}+
$$
\end{theorem}

\begin{theorem}
For large enough $\beta$ there is a single eigenvalue which asymptotically behaves as
$$
  E_1(\beta) = \frac{1}{2} - \frac{4}{\beta^4} + \mathcal{O}\left(\beta^{-5}\right).
$$
\end{theorem}

\section{Another model}
\label{sec:another}

The Smilansky-Solomyak model is not the only system in which the effect of an abrupt spectral transition can be observed. Now we are going to describe another model in which the transition is even more dramatic as a switch from a \emph{purely discrete spectrum} in the subcritical case to the whole real line in the supercritical one. To begin, recall that there are situations where \emph{Weyl's law fails} and the spectrum is discrete even if the classically allowed phase-space volume is infinite. A classical example of such a situation is due to \cite{S83} a two-dimensional Schr\"odinger operator with the potential $V(x,y) = x^2y^2$, or more generally, $V(x,y) = |xy|^p$ with $p\ge 1$. Similar behavior one can also observe for Dirichlet Laplacians in \emph{regions with hyperbolic cusps} -- see \cite{GW11} for more recent results and a survey; recall also that using the \emph{dimensional-reduction technique} of Laptev and Weidl \cite{LW00} one can prove tight spectral estimates for such operators.

A common feature of these models is that the particle motion is confined into \emph{channels narrowing towards infinity}; the increasing `steepness' of those `walleyes' is responsible for the discreteness of the spectrum. This may remain true even for Schr\"odinger operators whose potential are \emph{unbounded from below} in which a classical particle can escape to infinity with an increasing velocity. The situation changes, however, if the
\emph{attraction is strong enough}; recall that a similar behavior was noted already in \cite{Zn98}. As an illustration, let us thus analyze the following class of operators on $L^2(\R^2)$,
\begin{equation} \label{trapHam}
L_p(\lambda)\,:\; L_p(\lambda)\psi= -\Delta\psi + \left( |xy|^p - \lambda (x^2+y^2)^{p/(p+2)} \right)\psi, \quad p\ge 1,
\end{equation}
where $(x,y)$ are the standard Cartesian coordinates in $\R^2$ and the parameter $\lambda$ in the second term of the potential is non-negative; unless its value is important we write it simply as $L_p$. Note that $\frac{2p}{p+2}<2$ so the operator is e.s.a. on $C_0^\infty(\R^2)$ by Faris-Lavine theorem mentioned above; the symbol $L_p$ or $L_p(\lambda)$ will always mean its closure. Needless to say, the power in the last term is chosen in a way that makes it possible to play with the balance between the repulsion coming from the narrowing channels and attraction coming from the negative potential part.

\begin{figure}
\begin{center}
\includegraphics[angle=0,trim=0cm 11.5cm 10cm 11cm,width=8cm]{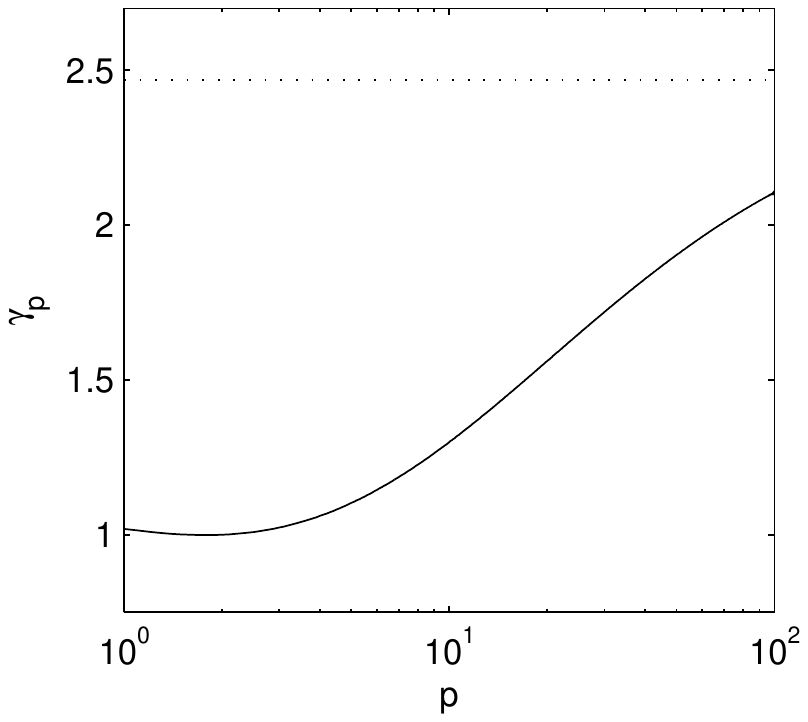}
\caption{$\gamma_p$ as a function of $p$ in the semilogarithmic scale}
\label{fig:6}
\end{center}
\end{figure}

Let us start with the \emph{subcritical case} which occurs for sufficiently small values of $\lambda$. To characterize the smallness quantitatively we need an auxiliary operator which will be an (an)harmonic oscillator Hamiltonian on line,
\begin{equation} \label{anharm}
\tilde H_p\,:\: \tilde H_p u = -u''+|t|^p u
\end{equation}
on $L^2(\R)$ with the standard domain. The principal eigenvalue $\gamma_p = \inf \sigma(H_p)$ equals one for $p=2$; for $p\to\infty$ it becomes $\gamma_\infty = \frac14 \pi^2$; it smoothly interpolates between the two values; a numerical solution gives true minimum $\gamma_p\approx 0.998995$
attained at $p\approx 1.788$; in the semilogarithmic scale the plot of $\gamma_p$ looks as shown in Fig.~\ref{fig:6}

As we have said, the spectrum is bounded from below and discrete if $\lambda=0$; our first claim \cite{EB12} is that this remains to be the case
provided $\lambda$ is small enough.

 \begin{theorem}
For any $\lambda\in [0,\lambda_\mathrm{crit}]$, where $\lambda_\mathrm{crit} := \gamma_p$, the operator $L_p(\lambda)$ is bounded from below for any $p\ge 1$; if $\lambda<\gamma_p$ its spectrum is purely discrete.
 \end{theorem}
\emph{Sketch of the proof}.
Let $\lambda<\gamma_p$. By the minimax principle \cite[Sec.~XIII.1]{RS} we need to estimate $L_p$ from below by a s-a operator with a purely discrete spectrum. To construct it we employ bracketing imposing additional Neumann conditions at concentric circles of radii $n=1,2,\dots\,$. In the estimating operators the variables decouple asymptotically and the spectral behaviour is determined by their angular parts ; to prove the discreteness one has to check that the lowest ev's in the annuli tend to infinity as $n\to\infty$. For $\lambda=\gamma_p$ this is no longer true but the sequence remains bounded from below.
\qed

A similar argument can be used in the \emph{supercritical case} with a few differences:
\begin{itemize}
 \setlength{\itemsep}{0pt}

 \item now we seek an \emph{upper} bound to $L_p(\lambda)$ by a below unbounded operator, hence we impose \emph{Dirichlet} conditions on concentric circles,

 \item the estimating operators have now nonzero contributions from the radial part, however, those are bounded by $\pi^2$ independently of $n$,

 \item the negative $\lambda$-dependent term now outweights the anharmonic oscillator part so that for the annuli operators $L^\mathrm{D}_{n,p}$ we have $\inf\sigma(L^\mathrm{D}_{n,p})\to -\infty$ as $n\to\infty$.

\end{itemize}
This yields the following conclusion \cite{EB12}:
 \begin{proposition}
 The spectrum of $L_p(\lambda),\: p\ge 1\,$, is unbounded below from if $\lambda > \lambda_\mathrm{crit}$.
 \end{proposition}

One can prove a stronger result, however, using a suitable Weyl sequences constructed in a way similar to that employed in the proof of Theorem~\ref{thm 2} it is possible to make the following conclusion \cite{BEKT16}:

 \begin{theorem}
 $\sigma(L_p(\lambda))=\R$ holds for any $\lambda > \gamma_p$ and $p>1$.
 \end{theorem}

In the subcritical case one can derive various results concerning properties of the discrete spectrum. Let us first mention an inequality obtain in a variational way for the proof of which we refer to \cite{EB12}. We define $\alpha:= \frac{1}{2}\left(1+\sqrt{5}\right)^2 \approx 5.236 > \gamma_p^{-1}$ and denote by $\{\lambda_{j,p}\}_{j=1}^\infty$ the eigenvalues of $L_p(\lambda)$ arranged in the ascending order, then we can make the following claim:

 \begin{proposition} \label{p:sumbound}
To any nonnegative $\lambda<\alpha^{-1}\approx 0.19$ there exists a positive constant $C_p$ depending on $p$ only such that the following estimate is valid,
\begin{equation} \label{ev_sum_bound}
\sum_{j=1}^N\lambda_{j,p}\geq
C_p(1-\alpha\lambda)\frac{N^{(2p+1)/(p+1)}}
{(\ln^pN+1)^{1/(p+1)}}-c\lambda\,N,\quad
N=1,2,\ldots,
\end{equation}
where $c=2\big(\frac{\alpha^2}{4}+1\big)\approx 15.7$.
 \end{proposition}

A similar, and even simpler result can be derived for regions with four hyperbolic `horns' such as $D=\{(x,y)\in\R^2:\: |xy|\le1\}$ which can be formally viewed as the limit of $p\to\infty$ of our model, and more rigorously they are described by the
Schr\"odinger operator
\begin{equation} \label{hypcusp}
H_D(\lambda):\: H_D(\lambda)\psi
=-\Delta\psi-\lambda(x^2+y^2)\psi
\end{equation}
with a parameter  $\lambda\ge 0$ and Dirichlet condition on the boundary $\partial D$. Following \cite{EB12}, one can make the following claim:

 \begin{theorem}
The spectrum of $H_D(\lambda)$ is discrete for any $\lambda\in
[0,1)$ and the spectral estimate
\begin{equation} \label{hypcusp_bound}
\sum_{j=1}^N\lambda_j\geq\,C(1-\lambda)\frac{N^2}{1+\ln\,N}\,\quad
N=1,2,\ldots,
\end{equation}
holds true with a positive constant $C$.
 \end{theorem}
\emph{Sketch of the proof}.
One can check that for any $u\in H^1$ satisfying the condition $\left.u\right|_{\partial D}=0$ the inequality
\begin{equation} \label{JMSest}
\int_D(x^2+y^2)u^2(x,y)\,\D x\,\D y\le
\int_D\left|\left(\nabla\,u\right)(x,y)\right|^2\,\D x\,\D y
\end{equation}
is valid which in turn implies $H_D(\lambda)\ge-(1-\lambda)\Delta_D$ where $\Delta_D$ is the Dirichlet Laplacian on the region $D$. The result then follows from the eigenvalue estimates on $\Delta_D$ known from \cite{JMS92, S83}. \qed

We we will not sketch the proof of Proposition~\ref{p:sumbound} because we are able demonstrate a substantially stronger result of Lieb-Thirring type \cite{BEKT16}:

 \begin{theorem} \label{thm:LT}
Given $\lambda<\gamma_p$, let $\lambda_1<\lambda_2\le\lambda_3\le\cdots$ be eigenvalues of $L_p(\lambda)$. Then for $\Lambda\ge 0$ and $\sigma\ge3/2$ the following inequality is valid,
\begin{eqnarray} \label{LT_bound}
\lefteqn{\mathrm{tr} \left(\Lambda-L_p(\lambda)\right)_+^\sigma} \\ && \le C_{p,
\sigma}\bigg(\frac{\Lambda^{\sigma+(p+1)/p}}{(\gamma_p-\lambda)^{\sigma+(p+1)/p}}
\ln\left(\frac{\Lambda}{\gamma_p-\lambda}\right)
+\,C_\lambda^2\left(\Lambda+C_\lambda^{2p/(p+2)}\right)^{\sigma+1}\bigg), \nonumber
\end{eqnarray}
where the constant $C_{p,\sigma}$  depends on $p$ and $\sigma$ only and
\begin{equation} \label{LT_const}
C_\lambda=:\max\left\{\frac{1}{(\gamma_p-\lambda)^{(p+2)/(p(p+1))}},\,\frac{1}{(\gamma_p-\lambda)^{(p+2)^2/(4p(p+1))}}\right\}.
\end{equation}
 \end{theorem}
\emph{Sketch of the proof}.
By minimax principle we can estimate $L_p(\lambda)$ from below by a self-adjoint operator with a purely discrete negative spectrum and derive a bound to the momenta of the latter. We split the plane $\R^2$ again, now in what one could call a `lego' fashion, cf.~Fig~\ref{fig:8}, using a monotone sequence $\{\alpha_n\}_{n=1}^\infty$ such that $\alpha_n\to\infty$ and $\alpha_{n+1}-\alpha_n\to 0$ holds as $n\to\infty$.
 \begin{figure}
 \label{fig:8}
 \setlength\unitlength{1mm}
 \begin{picture}(95,0)(30,72)
 \linethickness{.5pt}
 \put(56,70){\line(1,0){68}}
 \put(60,65){\line(1,0){60}}
 \put(60,15){\line(1,0){60}}
 \put(56,10){\line(1,0){68}}
 \put(60,10){\line(0,1){60}}
 \put(120,10){\line(0,1){60}}
 \put(65,15){\line(0,1){50}}
 \put(115,15){\line(0,1){50}}
 \put(65,55){\line(1,0){50}}
 \put(65,25){\line(1,0){50}}
 \put(75,25){\line(0,1){30}}
 \put(105,25){\line(0,1){30}}
 \put(56,74){\line(1,0){68}}
 \put(56,6){\line(1,0){68}}
 \put(56,6){\line(0,1){68}}
 \put(124,6){\line(0,10){68}}
 \put(90,60){$G_1$}
 \put(90,66.5){$G_2$}
 \put(90,71){$G_3$}
 \put(107,40){$Q_1$}
 \put(114.8,40){$Q_2$}
 \put(119.8,40){$Q_3$}
 \put(96,2){$x=\quad\;\alpha_1$}
 \put(114,2){$\alpha_2$}
 \put(118,2){$\alpha_3$}
 \put(123,2){$\dots$}
 \end{picture}
 \vspace{25em}
 \caption{Bracketing in the proof of Theorem~\ref{thm:LT}.}
 \end{figure}
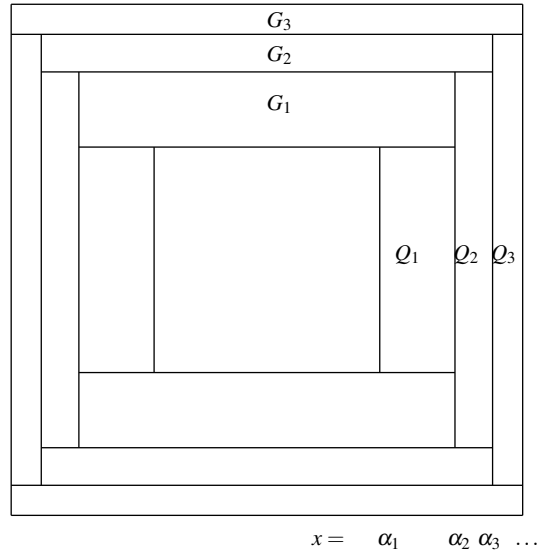
Estimating the `transverse' variables by by their extremal values, we reduce the problem essentially to assessment of the spectral threshold of the anharmonic oscillator with Neumann cuts. We derive easily the following asymptotic result:
\begin{lemma} \label{l:cut-osc}
Let  $l_{k, p}=-\frac{\mathrm{d}^2}{\mathrm{d}x^2}+|x|^p$ be the Neumann operator on $[-k, k],\:k>0$. Then
$$
\inf\,\sigma\left(l_{k,p}\right)\ge\gamma_p+o\big(k^{-p/2}\big)\quad\text{as}\;\;
k\to\infty.
$$
\end{lemma}

\medskip

\noindent In fact the error is exponentially small, but the above relation is sufficient for our purposes. Combining it with the `transverse' eigenvalues $\left\{\frac{\pi^2 k^2}{(\alpha_{n+1}-\alpha_n)^2}\right\}_{k=0}^\infty$, using Lieb-Thirring inequality for this situation \cite{Mi16}, and choosing properly the sequence $\{\alpha_n\}_{n=1}^\infty$, cf.~\cite{BEKT16}, we are able to prove the claim.
\qed

Let us finally look at the \emph{critical case}, $L:=-\Delta+|xy|^p-\gamma_p(x^2+y^2)^{p/(p+2)}$. The essential spectrum is as expected \cite{BEKT16} as one can check easily using properly chosen Weyl sequences:
 \begin{theorem}
 We have $\sigma_\mathrm{ess}(L)\supset [0,\infty)$.
 \end{theorem}

The question about the negative spectrum is more involved. First of all, we have the following result \cite{BEKT16} which can be proved by the same technique as Theorem~\ref{thm:LT} using another `lego bracketing' estimate:
 \begin{theorem}
 The negative spectrum of $L$ is discrete.
 \end{theorem}
For the moment, however, we are unable to prove that $\sigma_\mathrm{disc}(L)$ is nonempty. We conjecture that it is the case having a \emph{strong numerical evidence} for that. For simplicity consider the case $p=2$. We restrict the operator to a circle of radius $R$ with Dirichlet or Neumann boundary and compute the first two eigenvalues in both cases; they are plotted in Fig.~\ref{fig:8} as functions of the cut-off radius. By 
\begin{figure}
\vspace{-1.2em}
\begin{center}
\includegraphics[angle=0,width=7cm]{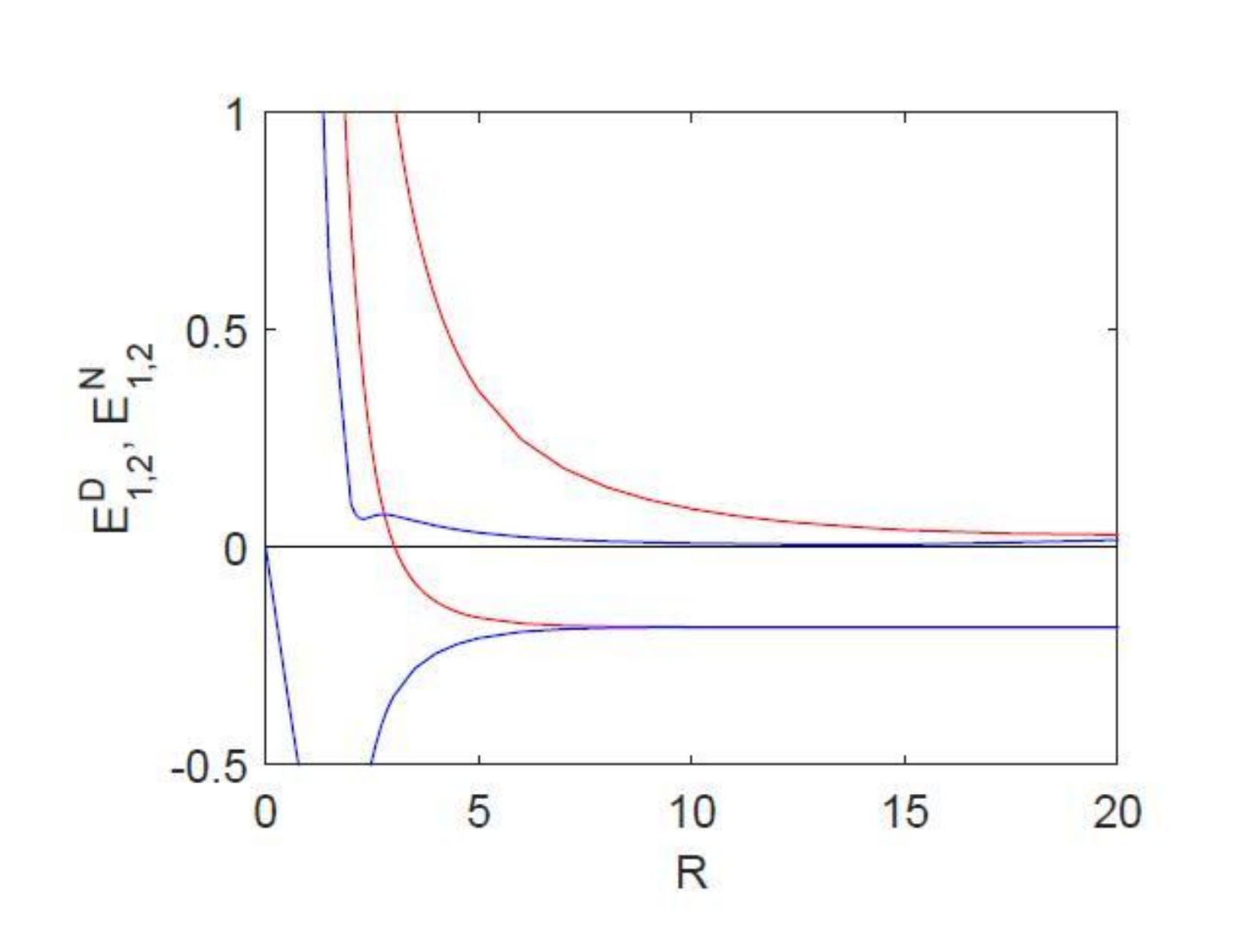}
\caption{The Dirichlet-Neumann estimate of the spectrum in the critical case for $p=2$.}
\label{fig:8}
\end{center}
\end{figure}
\cite[Sec.~XIII.15]{RS} the possible negative eigenvalues are squeezed between those curves. We see that the bounds become very tight for $R\gtrsim 7$ and indicate the critical problem has for $p=2$ an eigenvalue $E_1\approx -0.18365$. Furthermore, $\sigma_\mathrm{disc}(L)$ consists of a single point because the second lower (Neumann) estimate is in positive values for $R$ large enough. A similar numerical analysis also suggests the ground state existence for \emph{other values of $p$} but it becomes unreliable for $p\gtrsim 20$. We \emph{conjecture} that the \emph{discrete spectrum is nonvoid for all $p>1$ but empty for hyperbolic regions, $p=\infty$}. 

\begin{figure}
\vspace{-.5em}
\begin{center}
\includegraphics[angle=0,trim=0cm 1cm 0cm 1cm,width=9cm]{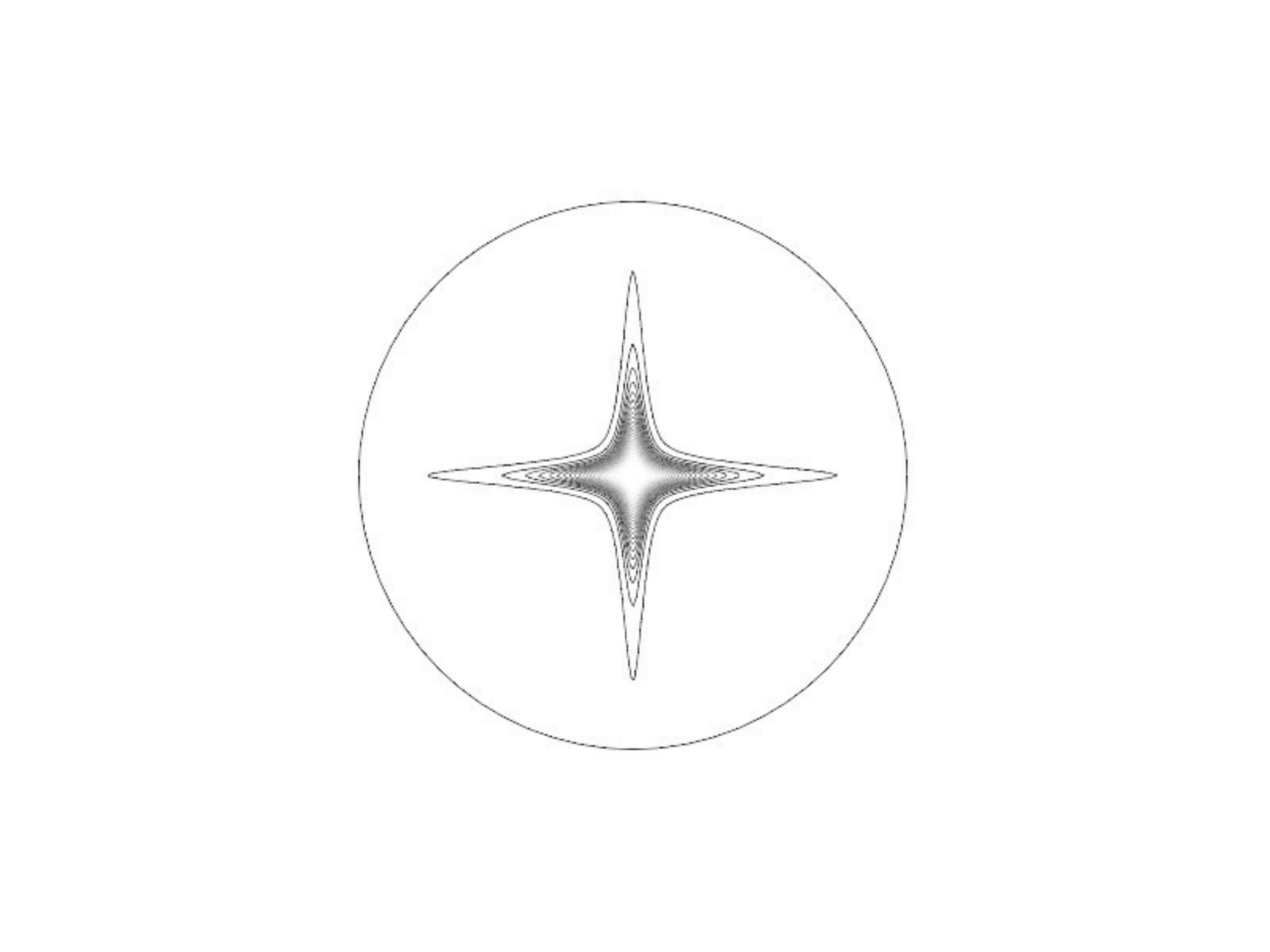}
\caption{The ground state eigenfunction in the critical case for $p=2$.}
\label{fig:9}
\end{center}
\end{figure}

We are able to get in a numerical way an idea about the ground state eigenfunction, again in the case $p=2$, as plotted in Fig.~\ref{fig:9} based on  
solution in the circle with either boundary condition; we note that with the $R=20$ cut-off the Dirichlet and Neumann ones are practically identical which is not surprising since one expects a superexponential decay along the axes. The outer level in the plot marks the  $10^{-3}$ value.

\section{Resonances in Smilansky-Solomyak model}
\label{sec:resonances}

The models we consider have other interesting properties. Let us return to the setting of Section~\ref{sec:smisom} and show that the system exhibits a rich \emph{resonance structure}; we refer to \cite{ELT17a, ELT17b} for a detailed discussion of these phenomena. To begin with, we have to say \emph{which resonances} we speak about. There are \emph{resolvent resonances} associated with poles in the analytic continuation of the resolvent over the cut(s) corresponding to the continuous spectrum, \emph{scattering resonances} identified with complex singularities of the scattering matrix.

The former are found using the same Jacobi matrix problem as before, of course, this time with a `complex energy'. Let is look at the latter. Suppose the incident wave comes in the $m$-th channel from the left. We use the Ansatz
\begin{equation} \label{scattAnsatz}
f(x,y) =\left\{ \begin{array}{l} \sum_{n=0}^\infty \Big( \delta_{mn} \mathrm{e}^{-ipx} \psi_n(y) + r_{mn}\, \mathrm{e}^{ix\sqrt{p^2+\epsilon_m-\epsilon_n}} \psi_n(y)\Big) \\[.5em] \sum_{n=0}^\infty t_{mn}\, \mathrm{e}^{-ix\sqrt{p^2+\epsilon_m-\epsilon_n}} \psi_n(y) \end{array} \right.
\end{equation}
for $\mp x>0$, respectively, where $\epsilon_n=n+\frac12$ and the incident wave energy is assumed to be $p^2+\epsilon_m =:k^2$. It is straightforward to compute from here the boundary values $f(0\pm,y)$ and $f'(0\pm,y)$. The continuity requirement at $x=0$ together with the orthonormality of the basis $\{\psi_n\}$ yields
\begin{equation} \label{rt}
t_{mn} = \delta_{mn} + r_{mn}.
\end{equation}
Furthermore, we substitute the boundary values coming from the Ansatz \eqref{scattAnsatz} into
\begin{equation} \label{scatt-bc}
f'(0+,y)-f'(0-,y)-\lambda yf(0,y) = 0
\end{equation}
and integrate the obtained expression with $\int \mathrm{d}y\, \psi_l(y)$. This yields
\begin{equation} \label{scatt-secular}
\sum_{n=0}^\infty \Big( 2p_n \delta_{ln} -i\lambda(\psi_l,y\psi_n) \Big) r_{mn} = i\lambda(\psi_l,y\psi_m),
\end{equation}
where we have denoted $p_n = p_n(k) := \sqrt{k^2-\epsilon_n}$. In particular, poles of the scattering matrix are associated with the kernel of the $\ell^2$ operator on the left-hand side. In particular, putting $l=m$ we obtain essentially the same condition we had before, cf.~\eqref{secular} and\eqref{Jacobi}, thus we arrive at the following conclusion:
\begin{proposition}
The resolvent and scattering resonances coincide in the Smilansky-Solomyak model.
\end{proposition}
Let us add a few comments:
\begin{itemize}
 \setlength{\itemsep}{0pt}
 \item The on-shell scattering matrix for the initial momentum $k$ is a $\nu\times\nu$ matrix where $\nu:=\big[k^2-\frac12\big]$ whose elements are the transmission and reflection amplitudes; they have common singularities.
 \item The resonance condition may have (and in fact it has) numerous solutions, but only those `not far from the physical sheet' are of interest.
 \item The Riemann surface of energy has infinite number of sheets determined by the choices \emph{branches of the square roots}. The interesting resonances on the $n$-th sheet are obtained by \emph{flipping sign of the first $n-1$ of them}.
\end{itemize}

\begin{figure}
\vspace{0em}
\begin{center}
\includegraphics[angle=0,width=10cm]{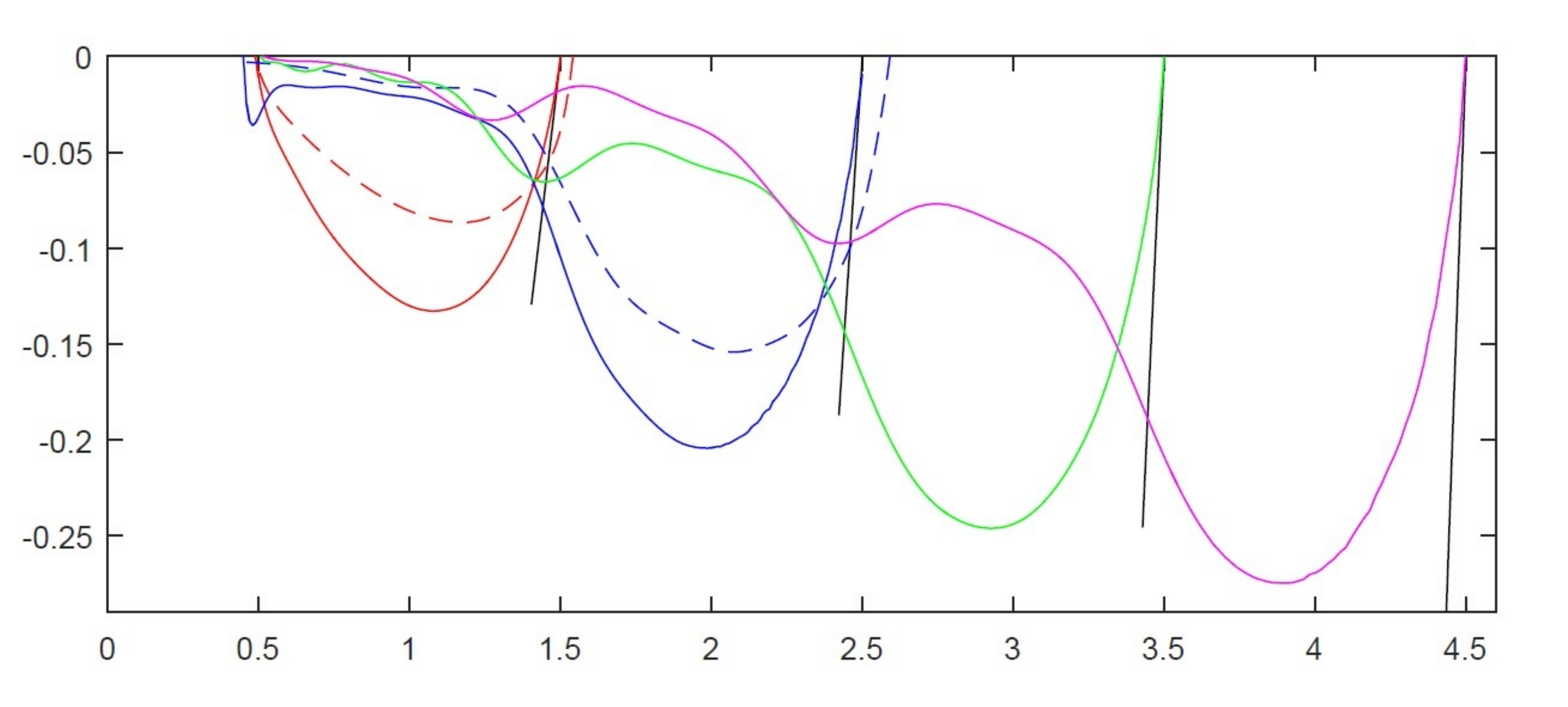}
\caption{Resonance trajectories as the coupling constant $\lambda$ varies from zero to $\sqrt{2}$.}
\label{fig:10}
\end{center}
\end{figure}

The \emph{weak-coupling analysis} follows the route as for the discrete spectrum, cf.~\eqref{component}--\eqref{Smilweak} above; in fact it includes the eigenvalue case if we stay on the `first' sheet. It shows that for small $\lambda$ a resonance poles splits of each threshold according to the asymptotic expansion 
\begin{equation} \label{weakres}
\mu_n(\lambda) = - \frac{\lambda^4}{64} \big( 2n+1 + 2in(n+1)\big) + o(\lambda^4).
\end{equation}
Hence the distance for the corresponding threshold is proportional to $\lambda^4$ and the trajectory asymptote is the `steeper' the larger $n$ is.
However, one solve the condition \eqref{scatt-secular} numerically \cite{ELT17a}. This allows us to go beyond the weak coupling regime and \emph{the picture becomes more intriguing} as shown in Fig.~\ref{fig:10}.
The picture shows clearly the asymptotes of the resonance trajectories for small values $\lambda$ when the poles split from the channel threshold given by the oscillator eigenvalues. For stronger coupling the behaviour changes and eventually the poles return to the real axis as $\lambda$ approaches the critical value. What is even more interesting, the numerical solutions reveals other, `non-threshold' resonances at the second and third Riemann sheet, indicated by dotted lines, that appear at $\lambda=1.287$ and $\lambda=1.19$, respectively.

\section{Concluding remarks}
\label{sec:concl}

While we have been able to demonstrate many properties of the models under consideration, various mathematical questions remain open, for instance
\begin{itemize}
 \setlength{\itemsep}{0pt}

\item in the original Smilansky-Solomyak model and its $\delta'$ modification of Sec.~\ref{subsec:delta'} we know that the essential spectrum is \emph{absolutely continuous}. We expect that this will also be the case for the models with regular potential channels but this remains to be demonstrated.

\item in the regular Smilansky-Solomyak model the `escape channel' may have more than one mode provided $\#\sigma_\mathrm{disc}(L)>1$ holds for the operator \eqref{comparison}. In this situation it is natural to ask how the \emph{spectral multiplicity} changes with $\lambda$.

\item many question concern \emph{resonances} in the Smilansky-Solomyak model. One would like to know, \emph{inter alia}, what is their number in a given part of the complex plane, whether there are resonance free zones for a fixed $\lambda$, or whether all the poles will eventually return to the real axis as $\lambda$ increases. Furthermore, we are interested in the mechanism which produces the `non-threshold' resonances and the coupling constant values at which they appear. Finally, resonance effects are also expected to occur in the regular version of the model.

\end{itemize}
From the physical point of view the most interesting question concerns the classical motion in the regular model, magnetic and nonmagnetic, as well as in the model of Sec.~\ref{sec:another}. We have mentioned in the opening of Sec.~\ref{subsec:regular} that a step in this direction was made in \cite{G18}, however, the importance of the question goes beyond the motivation of that paper dealing with modeling quantum measurements as it may offer a new and interesting insight into the quantum-classical correspondence in unusual situations we have discussed here.

\begin{acknowledgement}
Our recent results discussed in this survey are the result of a common work with a number of colleagues, in the first place Diana Barseghyan, Andrii Khrabustovskyi, Vladimir Lotoreichik, and Milo\v{s} Tater. whom I am grateful for the pleasure of collaboration. The research was supported by the Czech Science Foundation (GA\v{C}R) within the project 17-01706S.
\end{acknowledgement}
%

%
%
%

\end{document}